%% file: disco-arxiv.tex
\documentclass[5p]{elsart5p}

\usepackage[T1]{fontenc}
\usepackage[latin1]{inputenc}

\usepackage{url}
\usepackage{rotating}
\usepackage{multirow}
\usepackage{multicol}

\input{FigMacros}

\begin{document}

\begin{frontmatter}

\title{DISco: a Distributed Information Store for network Challenges and their Outcome}
\author[run]{Sylvain Martin \corauthref{cor}\ead{martin@run.montefiore.ulg.ac.be}}
\author[run]{Laurent Chiarello}
\author[run]{Guy Leduc}
\address[run]{Research Unit in Networking, Institut Montefiore, 4000 Liège, Belgium}
\corauth[cor]{corresponding author.}

\maketitle

\begin{abstract}
We present DISco, a storage and communication middleware designed to enable distributed and task-centric autonomic control of networks. 

DISco is designed to enable multi-agent identification of anomalous situations -- so-called ``challenges'' -- and assist coordinated remediation that maintains degraded -- but acceptable -- service level, while keeping a track of the challenge evolution in order to enable human-assisted diagnosis of flaws in the network. We propose to use state-of-art peer-to-peer publish/subscribe and distributed storage as core building blocks for the DISco service.
\end{abstract}
\end{frontmatter}

\section{Introduction}

Network monitoring mostly follow a location-centric, hierarchical processing \cite{COPS} of information where most decisions are ultimately made by human operators. We argue that this model suffer three major limitations. 
First, the reaction of human operators is limited, especially when problems mostly become as complex as nowadays botnet-driven denial of service or worm propagation. 
Second, the hierarchical approach allows some local manager to automate some ``reflex'' reaction based on local information. Any decision that requires knowledge -- even summarised -- regarding a larger area has to be deferred to a higher level in the hierarchy. As a result, the top of the monitoring hierarchy becomes a strategic target to intentional attacks: if taken offline or overloaded, defence of the network becomes severely compromised.
Finally, although the causes of service degradation are numerous, the analysis of a challenging situation is performed by a single program (e.g. an IDS) that must take into account all their possible variations. 

In line with the Resilinet/$R^2D^2+DR$ strategy \cite{Sterbenz20101245}, we argue that the role Internet now plays in our society and the evolution of challenging events stems for a generalised ability for the network to self-defend by activating \emph{remediation} mechanisms (traffic filtering, rerouting, ...) that will sustain service in a degraded, but acceptable state.
The need for a more resilient Internet suggests decisions always originate from a local system, possibly even defining areas of the network as Self-Managed Cells \cite{smc-noms}.
A \emph{challenge}, in this context, is an event that impairs operation of the network and therefore threatens the quality, availability of the services it delivers. This definition includes malicious attacks, mis-configurations, accidental faults and operational overloads. 

\begin{figure*}[t]
\begin{center}
\includegraphics[width=.8\textwidth]{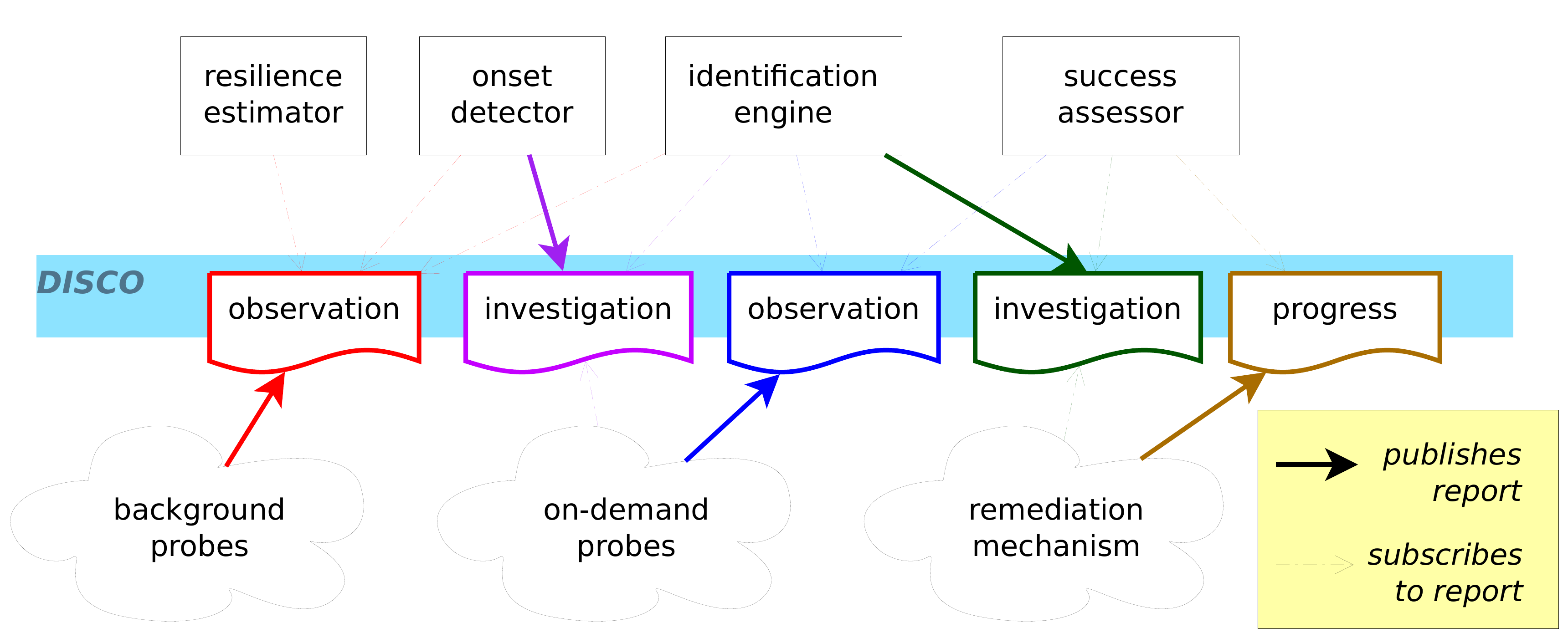}
\caption{Overview of the processing chain in detection and remediation of a challenge}\label{fig:disco-actors}
\end{center}
\end{figure*}

Appropriate autonomic response to network challenges stems for \emph{real-time} onset detection that acts as a background task, and triggers on-demand more sophisticated mechanisms involved in root-cause and impact analysis. This two-phase approach is crucial for saving resources of the network operator, as the (relatively) computationally expensive tasks can be focused both in time and in the amount of traffic they consider.

We argue that this multi-stage processing of monitoring information to ultimately provide a high-level understanding of a situation is similar to the process of hand-script recognition in architectural sketches \cite{esquise} and could therefore benefit from a multi-agent approach, where each agent is specialised (task-centric) into one kind of challenge and can identify it with a certain confidence level. Translating this approach to network control, however, requires a distributed middleware that can efficiently relay information and self-document its decision to allow human-driven revision of the policies. 

This paper collects the requirements for such a middleware (section \ref{sec:disco}), propose a service model (sections \ref{sec:disco:design}) -- the Distributed Information Store for Challenge and their Outcome -- that fulfil these requirements and review the available components it could be built on (section \ref{sec:disco:architecture}).
We further validate the concept on a DDoS detection scenario (section \ref{sec:disco:usage}) and evaluate the feasibility of using Scribe -- a DHT-based publish/subscribe service -- as the core building block (section \ref{sec:disco:impl}).

\section{Problem Statement}
\label{sec:disco}
We base our work on the findings of \cite{fry-ngi} regarding the detection of challenging situations in a network, that is, beyond \emph{onset challenge detection}, which is usually achieved through anomaly or signature-based detection, additional steps are required to \emph{classify} the challenge and understand its impact on the network and on ongoing communications. 
The ultimate goal of the challenge identification process is to activate and deploy a specific \emph{remediation mechanism}, providing it with all the required configuration parameters to handle the challenge and restore the service to a degraded, but acceptable level.
If we consider defence against DDoS attacks as an example, onset detection could consist of link and queues monitoring. On-demand identification involves a volume anomaly detection that pin-point the victim of the attack, and remediation mechanism would be a rate limiter for a firewall.

Figure \ref{fig:disco-actors} illustrates the communication chain involved in detection, identification and remediation of challenges through reports published into DISco. Probes present in forwarding plane components may vary in complexity, from single-variable monitoring (e.g. number of pending TCP connections) to more sophisticated entropy-based systems. 
These reports are gathered by onset detection agents and identification engines, usually from multiple sources, which in turn produce investigation reports when the rise or decay of a challenge happens.

Notification of challenges trigger the activation of mitigation components which will take actions to remediate the challenge. Because these decision are taken automatically and require a response time that no human operator can offer, it is important to record the progress of the challenge as a whole (triggering conditions, evolution of the impact, side-effects) so that the fitness of the automated solution can be analysed later to adjust response thresholds and parameters.

We identified three key problems that hinder the development and deployment of efficient detection and remediation techniques, and we suggest that a common information dispatching and storage system is the proper abstraction to address them:

\begin{itemize}
\item There may be many sensors, reporting more information than we can afford to relay on the network. Because we expect that multiple algorithms will be deployed, each operating as an autonomous agent to identify a specific kind of challenge, we want the sensors to remain unaware of the number and identity of their listeners. Moreover, the relative network location of detection algorithms and sensors impacts the accuracy we need on the information we receive.
\item Detection, remediation and diagnostic actions are delayed from sensing activities, yet they may require detailed information on past events that preceded a trigger. Therefore, the required lifetime of individual events is hard to predict, but it requires careful management given the potential amount of generated information.
\item New components will be deployed over time, to better identify and remediate unforeseen and foreseen challenges. They will likely alter the coupling between data by introducing new relationships and attributes. While this is an essential feature to guarantee successful evolution of machine-learning algorithms beyond their initial programming, it also implies that no database schema can be established in advance. Yet the dynamics of information makes identifier-based solutions such as IF-MAP \cite{www:IF-MAP} not applicable ``as is''.
\end{itemize}

To address these problems, the Distributed Information Store for Challenges and their Outcome (DISco) provides the following features:

\begin{itemize}
\item a aggregation-capable publish/subscribe function that relays information between sensors, detectors, and mitigators.
\item an annotation system, coupled with more conventional database-like lookups that allows detectors and mitigators to further classify sensors information and adjust its lifetime accordingly.
\item a distributed (peer-to-peer) storage system that provides system-wide longer-term persistence for data that have been ``elected for diagnostic'' taking into account the existence of ``natural'' storage space such as routing tables.
\end{itemize}

%

\section{Design Principles}
\label{sec:disco:design}

The following design principles steered us from the problem description to the architecture proposed in section \ref{sec:disco:architecture}.
\begin{description}
\item[Evolving System]: 
We expect that the monitored system will evolve by addition of new probes and agents over time. As a result, new type of information and new information processors will appear and must integrate the protected network without requiring to reprogram, reconfigure or reboot other components.

\item[Peer-to-Peer Distributed System]: 
Distributed monitoring infrastructures typically follow a strongly hierarchical approach where a device in a low level of the hierarchy receives and process an important amount of information about a small area and report its conclusions to the level immediately above, up to a central device that has coarse and complete view of the network and take decisions that are forwarded and enforced down the hierarchy. We argue that, although offering interesting locality properties, such an approach lacks scalability by the fact it excessively decouples data-plane monitoring and enforcement from decisions that are delegated to the management plane.

Advances in structured peer-to-peer hash tables and messaging (especially pub/sub) systems would comparatively allow any number of devices to cooperate so that the \emph{control} plane of a device can obtain network-wide context it lacks to locally process fine-grained events describing its own behaviour, decide the required changes and inform peers of its decision to avoid inconsistent global behaviours.

\item[Multi-Resolution Information]: 
When an agent receives an observation report, it is essential that it can gather additional information that was not explicitly included in that report. Through the coupling between the pub-sub system and the distributed database, agents are allowed to ``zoom'' into an event by collecting additional information with specified scope in time, location and layers. This is a principle we share with \cite{Yang09FlowNet}, although the rest of our approach differs from that proposal.

\item[Learning-Ready Data Model]: 
Information relayed by DISco must ultimately be useful as input to machine learning algorithms, that will provide configuration parameter of the adaptive probes and remediation mechanism, but also to automate the identification of meaningful symptoms for a given problem among a huge amount of measurable parameters. 
To that regard, we preferred machine-oriented representation (tuples of numbers) over log entries. Alarms, reports, notification are mapped to such tuples where a special member serves the purpose of identifying the \emph{nature} of the event (the event identifier) and the rest consists of an arbitrary amount of attributes.

\label{sec:disco:voc}
Identifiers for events and attributes relayed through the DISco need to have a commonly agreed semantic for all components in the system. We propose to reuse for this purpose the concept of \emph{Vocabulary Specification Trees} (VSTs) described in the monitoring framework of the European ANA project \cite{ana-d37}. Vocabularies provide easy-to-manage and extensible collection of terms that are hierarchically organised by a ``IS-A'' relationship. We would, for instance, express that \texttt{bandwidth} is a connection-related metric by placing it under the \texttt{connection} node of the vocabulary tree rooted at \texttt{metrics}. As with ontologies, VSTs allow to distinguish concepts: \texttt{metrics.connection.bandwidth} is different from \texttt{resources.link.bandwidth}, which express how fast raw data can be sent over a link. Each concept is thus a node in a tree and can be refined by adding children concepts to it.

\item[Keeping Management Apart]: 
Tasks occurring in the management plane typically occur at a different pace and with a level of abstraction that differs from control and data planes. Therefore, our proposal doesn't include any mechanism for, e.g., (human-readable) self-description of exchanged information. Assignment of numerical identifiers to event and attributes, for instance, can be synchronised independently with assignment of other devices at initialisation. Numerical IDs are suited to run-time processing, while their connection to concepts in the VST allows extension to a richer database, e.g., by linking reports to metrics as ``\texttt{challenge.X} --\emph{impairs}$\rightarrow$ \texttt{metrics.Y}''. This database can be stored in a separate relation table available to human-assisted, or solver-based, diagnostic and refinement tasks.
\end{description}

\section{Architecture and Main Components}
\label{sec:disco:architecture}

As we want DISco to serve both for dissemination and storage of information (more specifically, \emph{events} reports), it will indeed consist of a peer-to-peer publish \& subscribe function, combined to a distributed storage to provide data persistence at short and longer-term. A general view of this architecture is depicted on Figure~\ref{fig:DISco-architecture}.

\fig[1]{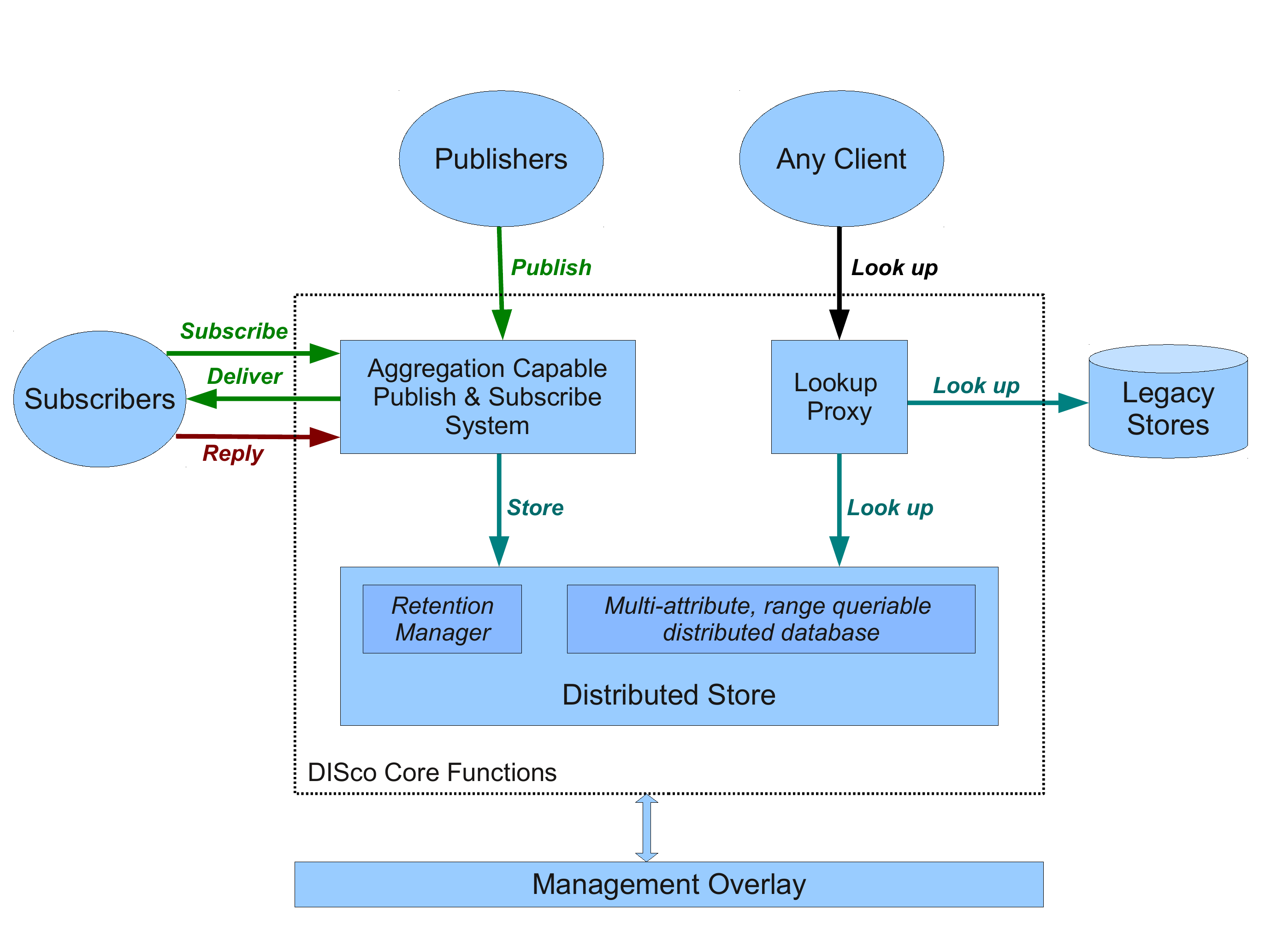}{General architecture of DISco.}

We collectively name \emph{clients} the software components (or ``blocks'') that make use of the DISco API, both for delivery of events or searches through the storage facility. \emph{Publisher} and \emph{subscribers} are merely roles, and it is frequent that a single component play both for different levels of event. Every client -- detectors, identifiers, remediators and off-line diagnostic tool -- has access to data archived in the store through the \emph{lookup} interface, which enables among other things on-demand resolution ``zooming''. The content of this store can be conceptually extended to legacy storage that could provide useful context, such as routing tables, as described in Section \ref{sec:disco:architecture:storage}.

\subsection{Publish \& Subscribe System}

The Publish \& Subscribe mechanism is used for near real-time notification of events, and is thus the privileged path for information exchange between detection and remediation agents. While detection agents are quite naturally associated to publishers, and remediation agents to subscribers, it is also expected that several \emph{identification engines} will be using the information  extracted from various sources they subscribed to, in order to publish some higher-level events.

Beside these two actions (\emph{publish}, \emph{subscribe}), DISco will be providing a \emph{reply} mechanism, that can be used to annotate the data. A subscriber can then tell the notifier that the published summary of some data is of importance and that the detailed data should be kept for a longer duration. The annotation tags can provide link between low-level and high-level (correlated) events, trigger the storage of these related events in the distributed store, and might also guide the auto-configuration of aggregators.

\subsection{Filtering and Aggregation}
\label{sec:disco:architecture:aggregation}

DISco is intended to be deployed in large systems, with possibly huge volume of information to process. While there is no \emph{a priori} limitations to the amount of published data and granularity of events, a practical solution has to keep bandwidth and storage usage as low as possible. This leads to two possible solutions: either publishers are responsible for limiting the amount of events they generate, or the subscribers specify to the system the granularity level they are interested in.
However, only the second approach will succeed in efficiently reducing the data volume without losing essential information, mainly for two reasons. First, publishers have no (or only few) knowledge of the level of interest in their publications, possibly leading them to largely under- (or over-)estimate the optimal granularity. Second, subscribers will be able to dynamically reconfigure (e.g., through a new subscription) the desired granularity in order to tell publishers to be more verbose when they detect suspicious events. 
This allows to maintain relatively low management overhead during normal operations, while gathering more precise data during challenges, enabling better detection and/or diagnosis. %

When subscribing to a particular event, the subscriber can specify limitations to the amount of published notifications through
\begin{itemize}
 \item \textbf{Filters}, discarding non relevant events and/or attributes, and
 \item \textbf{Aggregators}, gathering several events to produce a single, coarser-grained notification.
\end{itemize}

It is important to underline that the only role of aggregation is to merge similar events, the combination of which is an event of the same type (attributes are altered, though). This must not be confused with correlation, which extracts information from several events in order to deduce one of a new type. This correlation is not performed by the DISco itself, but in clients such as \emph{identification engines}.

When specifying aggregation, the system has to be told \emph{how} and \emph{how much} to aggregate. Although it could be imagined to let full freedom in this specification, through, for instance, programmable aggregators, this would raise huge implementation (and even maybe security) problems. Instead, predefined aggregators will be selectable, depending on the type of event, and following the subscribers' needs. 

Considering the granularity level, two kinds of specification can be provided. On one hand, the event rate has to be controlled strictly, and the aggregator delivers a periodic summary, containing more or less base events, depending on the number of generated events during the interval considered. On the other hand, we may require that a single aggregated event always contain the same number of base events. In this case, traffic will not be uniform over time, allowing to better capture critical situations.

\subsection{Distributed Storage System}

The published events are kept in a distributed store across participating nodes for further analysis. This includes both detection algorithms requesting recent events (short-term storage), and delayed processes running diagnosis on a larger scale (long-term storage).

Many distributed stores are based on \emph{distributed hash-tables} (DHT), basically using a hash of an element identifier to determine the node on which it has to be stored. While this approach is used by many peer-to-peer systems which need to search single elements based on their identifiers, DISco is required to handle more evolved lookups, supporting range-based queries on several attributes (such as IP range, time intervals, thresholds on values, and so on), while considering others as wild-cards.

Other structures have been developed to handle this kind of queries, but we identified only two of them being of interest and supporting \emph{multi-attribute} range queries: \emph{Mercury} \cite{mercury} and \emph{SkipTree} \cite{skiptree}. The latter has been selected for implementation and use in DISco, since it has two major advantages over Mercury: locality properties and lower space usage. %

\subsection{Information Retention}

Ideally, element removal will be something that is handled in an autonomous way by DISco itself through the \emph{Retention Manager}. It will use information such as number of subscribers, past lookups, and annotations as hints that a specific data entry needs to be ``promoted'' to a longer storage (typically, for the diagnosis phase). Static and manual configuration should be used only to define defaults and characterise retention length depending on the hints mentioned above.

\subsection{Heterogeneous Storage}
\label{sec:disco:architecture:storage}

While DISco provides, from a logical point of view, a single store (even if physically distributed) for all published events, we suggest to organise storage facilities in three distinct classes.

\emph{Local Temporary Storage} (LTS) is co-located with sensors and stores for a short amount of time a copy of every published event, regardless of its potential value and the existence of subscribers. The persistence of LTS is usually limited to a small multiple of edge-to-edge domain latency as it is solely intended to give subscriber the opportunity to \emph{reply} to an event.

\emph{Distributed Working Storage} (DWS) is the main storage facility, based on SkipTree (or a similar alternative), that holds and organises pertinent published information from onset detection to off-line diagnostic.  

Finally, \emph{Legacy Storage} (LS) consists of pre-existing, ``natural'' stores of information (such as BGP routing tables) that are made accessible through an additional translation daemon running on their host. The content of those legacy stores could indeed be a valuable source of context for many algorithms, and we could benefit on having a unified way to reach them.
Instead of duplicating legacy information directly in the DWS, only location hints would be kept, relying on the \emph{lookup proxy} to follow indirections transparently. This daemon-indirection-proxy chain of components should be sufficient for further extension needs.
Context present in the LS can be selectively transferred to the DWS with a \emph{reply} on the lookup result in order to capture relevant data for later diagnostic.

\subsection{Connectivity Layer}

For proper operation, DISco needs a resilient communication infrastructure that can be provisioned with limited, but guaranteed bandwidth (i.e., defended against volume-based attacks), and possibly using secure channels (i.e., encrypted communications and known partners).
Its goal is to decouple DISco-related traffic (i.e., management traffic) from the monitored traffic. We also assume that it provides peers authentication and integrity of the DISco-internal traffic so that neither forgery nor falsification of reports could occur. Achieving this level of resilience is beyond the scope of this paper. We will simply assume that appropriate ``detour tunnels'' exists, that prevents \emph{common} challenges from impairing connectivity of DISco nodes.

\section{Concept Validation : DDoS Detection}

\label{sec:disco:usage}

\subsection{Network-Network Interaction}

To illustrate the way DISco works, let us use the following denial of service attempt %
on the network depicted in Fig. \ref{fig:disco-drops}. Attackers target link $L$ that is required to reach a victim attached to $V$. They additionally identified that traffic towards destinations attached to $U$ also uses this link and, thus, uses addresses $U_i$ as well to dilute the signature of their attack. We also assume that attackers decided to have their attack traffic dropped a few hops after $L$\footnote{by means of their Time To Live field or any similar hop count limiting technique} in order to further evade detection by systems in $V$'s network. This, however, makes their traffic look singular in the network containing $L$.
\fig[.75]{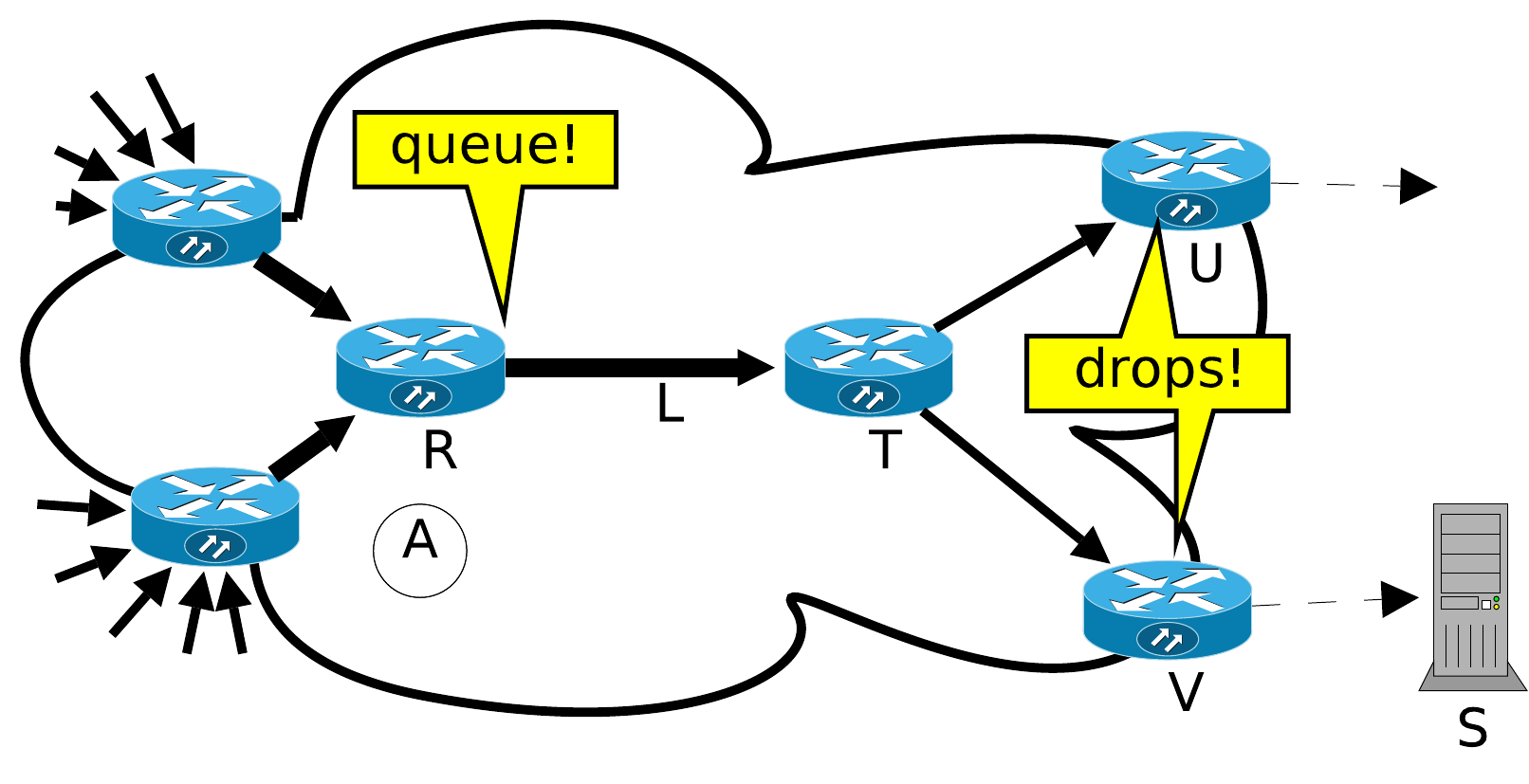}{Network under DDoS challenge, $L$ being is the overloaded bottleneck link, $R-T-V$ the path to the victim through the network}

Routers in this example monitor the amount of IP packets that are rejected by the forwarding process, including those who have their time to live exhausted. They report this through \texttt{event.network.drops.forwarding.\-rfc791-ttl-exceeded} which contains (as attributes) flow identification (made of source/destination addresses, ports and transport protocol, in the case of IPv4), location of the reporting router and timestamp of occurrence. These events are published by $V$ and $U$ in the distributed store. Similarly, an important amount of ``queue full'' events occurring at $R$ will trigger the execution of DoS-detection algorithms local to $R$ such as identifying destinations of largest flows. A further analysis process $A$ that was dormant in a system close to $R$ previously subscribed to ``any heavy-flow report event'' from $R$, and possibly other routers in the same point of presence (PoP).

Upon reception of heavy flows reports from $R$, $A$ will additionally subscribe to events reporting network-related errors downstream from $R$, enabling collection of reports from $T$, $V$ and $U$. We assume here that $A$ is an ``expert'' software agent that looks for and identifies the specific kind of DDoS attack we described above. The publish/subscribe mechanism fully allows multiple similar systems to execute concurrently and perform their own analysis using the same initial events.

The following features of DISco are highlighted in this example:
\begin{description}
\item[Local holding of data:] information about packets dropped at $V$ and $U$ are put under the control of DISco, but not yet transferred to a remote system until interest in such information is expressed through a \texttt{subscribe} call. Yet, it is important that such data can be looked up a posteriori, for instance when process $A$ tries and gathers recent past statistics to figure out the dynamics of the challenge. Temporal aggregation can still be applied to reduce the available granularity of information over time.

\item[Selective subscription:] while $A$ needs extra information from $T$, $U$ and $V$, it is only interested in information related to a fraction of the traffic those routers forward. For instance, if it identified 4.2.0.0/16 to be the destination of heavy hitters, we will add a constraint on attributes stating that \texttt{attribute.flow.rfc791-destination-address} must match that range.

\item[Compound values:] note that while $A$ describes filtering on ``destination address'', $V$ and $U$ put flow identification together in a compound value. The schema of this compound needs to be known by DISco peers so that filtering/aggregation components are able to extract and compare the destination addresses.

\item[Aggregating events from multiple sources:] $A$ typically makes no difference between reports coming from $U$ and $V$ as long as they match the filter. This highlights the need for describing a region of the network through an attribute constraint.

\item[Flexibility:] $A$ could broaden its monitoring criterion by subscribing to \texttt{event.network.drops*}, and receive notification of packet losses in the network regardless of whether they are due to TTL issues, congestion (queue full or early notifications), broken link, unknown destination, etc. This relieves $A$ from knowing the actual network protocol stack details (as a ICMP snooping agent would have to) and allows monitoring of events originated by different layers as needed.
\end{description}

\subsection{Network-Server Interaction}

We then consider a more ``classical'' DDoS attack, where the victim server $S$ is really receiving application-level requests through transport-level connections. $S$ locally observes those request patterns and their effect on system resources such as CPU and memory load, or access to internal databases. Deviation from sustainable behaviours are reported as events deriving from \texttt{event.server.overload.*} and include at least flow identification of the ``faulty'' connection.

When the agent $A$ coaching a router like $R$ observes abnormal traffic share towards $S$'s prefix, it may subscribe to ``server overload'' events to help deciding whether the currently observed challenge is a DDoS attempt. This, assuming that a DDoS is more likely to use resource-intensive requests while, during a flash crowd, time and resources needed to serve the requests do not deviate from normal behaviour and only the amount of requests per unit of time gets wild.

Similarly, resilience agent co-hosted with $S$ would subscribe to ``challenge detection reports'' and ``remediation action reports'' produced by networks delivering traffic to $S$. If needed, the DISco peers that receive this subscription can ensure that the agent only subscribe to information it is entitled to receive (that is, check the presence of an IP-destination filter).

This approach is especially attractive in content delivery networks (CDNs) \cite{conf/imc/KrishnanMSJKAG09}, where a single economic entity owns both the access network/routers and server farms. It can still be of high interest as a way to train learning-capable detectors. In that alternative, detection and remediation algorithms do not directly use ``server overload'' events, but instead use information coming from multiple symptoms to identify symptoms combination that reveal a DDoS challenge. In a refinement step, diagnostic agents look up for overload events in the time interval between ``challenge detected'' and ``end of challenge detected''. The correlation between the expected state (not challenged, detection in progress, remediation applied, ...) with the number of ``overload'' reports is used to assert the suitability of the detection process.

\section{Aggregation-Capable Pub/Sub over Scribe}
\label{sec:disco:impl}

This section focuses on the choices we made to implement the aggregation-capable pub/sub system required by DISco. The idea was to rely on the already-available implementation of Scribe \cite{scribe} for the OMNET++ simulator\footnote{\url{http://www.omnetpp.org/}}, included in Oversim\footnote{\url{http://www.oversim.org/}}~\cite{Oversim}. We then needed to adapt its mechanisms in order to integrate aggregation and filtering in its message delivery process. However, as explained with more details in Section~\ref{sec:disco:impl:inadequacy}, the mechanisms used by Scribe (and, to an extent, Key-based routing systems using DHTs) to deliver messages do not support the full range of features we require for our aggregation-capable system.

As described in the following text, Scribe proved to be a non-optimal choice for our purpose, the same way DHT wasn't ideal for the store itself. In parallel to the development of an alternative based on SkipTree, we decided to pursue the implementation of aggregation/filtering features over Scribe as a reference point.

\subsection{Adding Aggregation to Scribe}

Scribe uses Pastry~\cite{pastry}, a DHT system capable of routing messages to the node whose numeric ID is the closest to the message key. Using this property, each topic of the publish/subscribe system will be hashed to obtain a topic ID, and the node with the closest node ID will act as a rendez-vous point. During the subscription process, the message from the subscriber will be routed to the rendez-vous point. Each node on the path will subscribe to that topic and become a forwarder, until the message reaches a node being already a forwarder for that topic (or the rendez-vous point ultimately). This ends up with a multicast tree rooted at the rendez-vous point. Other mechanisms are used to maintain only active nodes in the tree (refresh), or even to change the root node in case of failure, but these will not be described here.

For our aggregation-capable system, each subscription

specifies aggregators in addition to the topic ID. The message is then routed towards the rendez-vous point as in the classical Scribe implementation. When a forwarding node in the multicast tree has children with different aggregation requirements,
its own subscription is aligned onto the finest-grained one. The necessary additional aggregation is performed on subsequent publications to serve children with coarser-grained requirements.

\subsection{Data Model, Templates and Discarded Attributes}

Since DISco is designed to be used by many devices and network components, it is essential to be able to make it evolve dynamically by adding new events and/or attributes progressively and smoothly. 
Moreover, since the vocabulary may be huge and not known by every DISco client, we need a mechanism that guarantees easy and consistent data formatting between clients. For this purpose, DISco uses an approach similar to IPFIX (\emph{IP Flow Information Export}~\cite{RFC5101}).

The format of the published events is described in specific \emph{Template} messages. These contain an ID (unique to the issuer) and the format description (attribute IDs and types) of the following events. Consequently, each published event will contain the ID of the corresponding template. It is worth noting that, due to aggregation and discarding of attributes, the template is likely to be different from node to node in the multicast tree.

When a subscriber receives a template for a particular subscription, it will be able to read all the attributes of the events, but may not be interested in all of them. In order to reduce bandwidth usage, DISco allows a subscriber to explicitly discard a list of attributes, preventing them from being sent subsequently.

\subsection{Generic Aggregators}

Based on the subscription details, nodes in the multicast tree will have to aggregate events and thus need functions to combine them. As already mentioned in Section~\ref{sec:disco:architecture:aggregation}, DISco does not provide fully programmable aggregation but instead, lets the subscriber select the aggregator amongst predefined ones.

The aggregation is applied on attributes individually. Each attribute receives an identifier through the vocabulary. Subscription defines the operation type to be applied, while event content defines the attribute type (this information is extracted from the template, preventing routers to cache knowledge of every \emph{attribute ID} $\rightarrow$ \emph{attribute type} relation). It is required that DISco has an aggregation function for each possible $<operator>$ x $<attrtype>$ pair. Core operators and attribute types are available, but new ones may be added to the system later on, provided appropriate aggregation function are deployed too.

\subsection{Basic Node Operations}

Several operations have to be accomplished by each node in the multicast tree in order to perform aggregation. Moreover, some state is required to maintain the list of children nodes, associated with their filtering and aggregation specification. This state is built up or updated as nodes subscribe to a particular topic. A published event goes through the chain of operations illustrated on Figure~\ref{fig:ulg-scribagg-deliver}.

\fig[0.8]{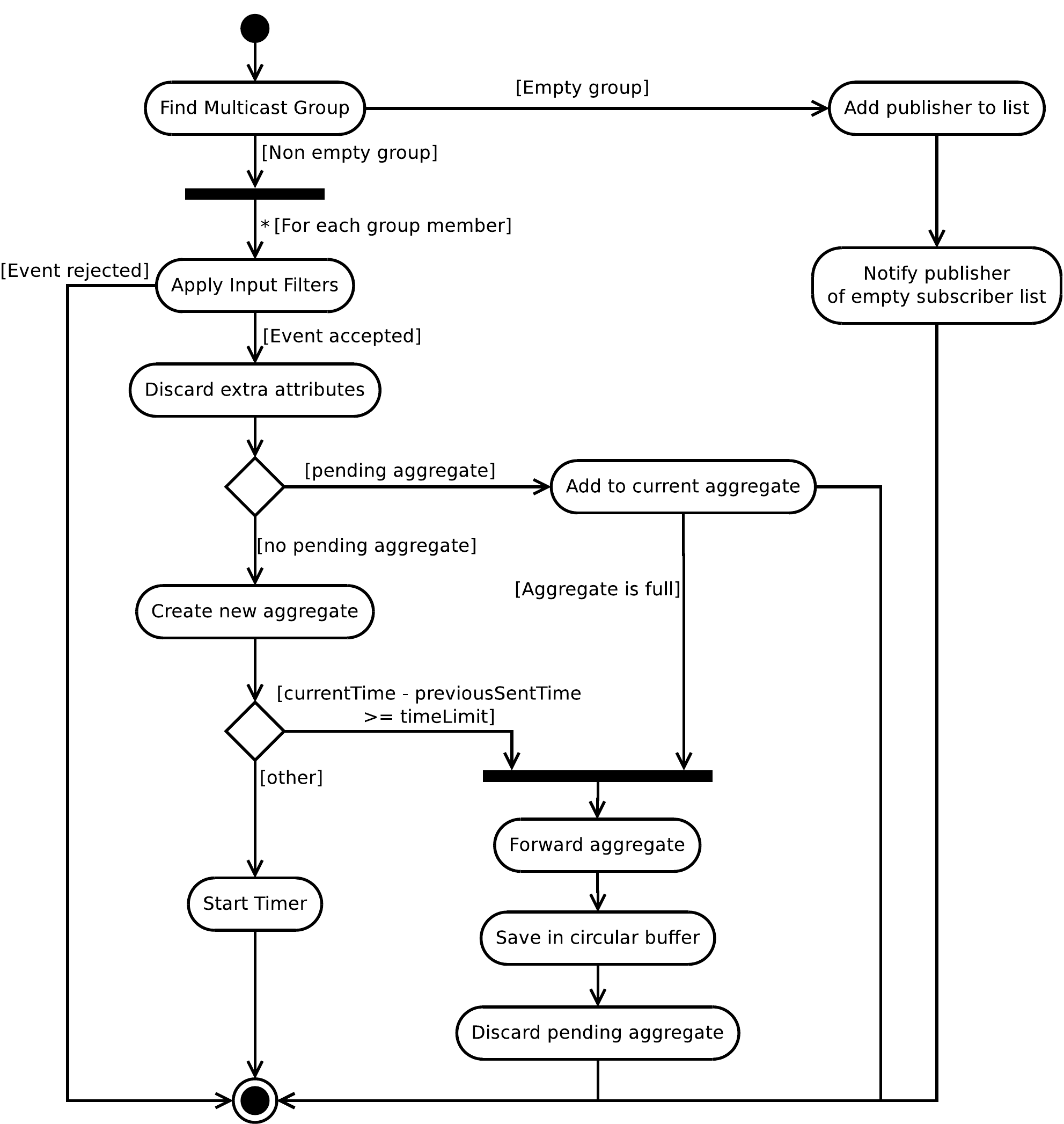}{Activity diagram of event aggregation and delivery by a DISco node.}

A DISco node maintains a list of each multicast group it belongs to, either being a subscriber (leaf node) or an internal node of the multicast tree. When it receives an event, the node must first of all find at which multicast group it is aimed. If no group is found, then the event was published before any subscription. It is worth noting that, unlike Scribe, publishing to a topic with no subscribers will be reported by the root node through a specific notification. This allows sources to regulate their publishing rate when there is no interest in their events. A list of early publishers is kept in order to notify them when subscribers are ready.

After this initial step, the event is filtered and aggregated following the specification of subscribers. The state of filters and aggregators for each direct children in the multicast tree  has to be maintained separately (since they may have diverse requirements). 

Since the level of aggregation can be specified through a maximum number of aggregated events and through a maximum period length, several situations have to be dealt with. The new event may be added to a pending aggregate, if any, or a new aggregate will be created. In the first case, we want to check if the maximum number of events is achieved and, in such a case, forward the aggregate. In the second case, we need to ensure that the maximum period is not already exceeded (meaning that the aggregate will be forwarded with only one event aggregated). After the creation of the new aggregate, we also need to start a timer that will trigger the forwarding at the end of the maximum period.

A local circular buffer is used for short-term storage of forwarded events. This buffer is used in case of replies to match the source events and further process them. The pending aggregate is then discarded and any associated timer cancelled.

\subsection{Scribe Inadequacy}
\label{sec:disco:impl:inadequacy}

\subsubsection{Multicast in presence of single publishers.}

When a specific topic is filled by only one publisher, the multicast tree should be rooted at the publishing node to avoid unnecessary traffic between the publishing node and the root of the multicast tree. But, since the rendez-vous point is chosen following the topic ID (a hash value), all the root nodes are pseudo-randomly distributed over all the nodes in the overlay. This is indeed a key point for load balancing in generic applications using Scribe. But, in our case, topics with only one publisher will typically correspond to precise, low-level events, whose subscribers are reasonably some local managers located quite near the publisher. Having to forward every single event through a possibly distant rendez-vous point would clearly increase the amount of traffic and impact on the minimal level of aggregation required to limit the overhead of management traffic.

\subsubsection{Topic aggregation through vocabulary.}

It is expected to have remediation agents subscribing both to specific events (e.g., \texttt{report.intrusion.protocol-\-exploits.rfc792.cve-\-1999-\-128}), because they know what they can remediate exactly, but admittedly a generic manager should be able to subscribe more largely (e.g., \texttt{report.intrusion.*}). Also, a remediation agent might want to be a bit less specific and subscribe to \texttt{.protocol-\-exploits.rfc792.*}. The DISco service model allows this to be addressed through the vocabulary, that is, ensure that all items under \texttt{report.intru\-sion.*} will be given IDs that match a common prefix, e.g. \verb+CA:FE/16+, while \texttt{alarm.failure.*} would be \verb+DE:AD/16+, much like IP addresses blocks.

Unfortunately, when using DHT-based routing for topics, the topic hierarchy is completely lost during the hashing operation, and topics can no longer be grouped. To work this limitation around, we use in our simulations an ``\emph{oracle}'' to determine which topics needs to be aggregated. Publishers and subscribers obtain the topic ID through this oracle instead of directly hashing the topic name. A single multicast tree will thus be built for several related topics, and internal nodes will be responsible for filtering out the events to be forwarded downward in the tree.

\subsubsection{Geographical aggregation.}

A third way of performing aggregation is to aggregate multiple sources based, typically, on their IP range. It would also be interesting to be able to aggregate events along a specified path, or in the vicinity of a particular device (both "path" and "vicinity" having to be defined). As for the previous drawback, there is no clear way of implementing these functionality when using key-based routing on a DHT, which purposely distribute similar keys in a pseudo-random manner.

\section{Extending Operation to Multi-domain Detection}

Initially, DISco is designed to federate information of multiple sources of event and made it available to entities that supervise network operation with both time and space multi-resolution. Although it can be applied at different scales (from campus network to e.g. the G\'eant research network or overlays), it assumes that all participants (probes, analysers, mitigators, resilience managers, ...) share a common interest for network resilience (and in most cases, a common administration), and that they can establish secure communication with each other.

As we raise towards the application-layer challenge detection and remediation, however, there is an increased need for inter-domain information exchanges. Decisions in autonomous domain $A$ may then be depending on (or influenced by) reports generated by probes in other autonomous domains. The major point to address is then to ensure that integrating those reports in our detection mechanism does not open the door to new attacks.

Additionally, DISco relies on a shared \emph{vocabulary} between running components, but doesn't require the vocabulary to be common among \emph{every} autonomous system. Inter-domain operation thus requires that either sessions first negotiate a vocabulary mapping, or that every exchanged message uses the textual representation of event and attribute identifiers, with implied overhead.

\subsection{Import/Export}

A conservative approach consists of programming explicit \emph{importers} and \emph{exporters} at the border of a DISco domain. An exporter subscribes to reports of its own domain, applies additional filtering and aggregation as defined by domain managers' confidentiality policies and relays resulting messages to the importer of a remote domain over a secure channel. The importer initially gather information about exporters' capabilities (such as IP ranges they report about) and informs the pub-sub system that it can potentially publish reports for the identified ranges. It also perform sanity filtering on received reports to avoid source spoofing of DISco message (i.e. it has to ensure that message coming from domain $D$ actually give information about domain $D$).

This import/export mechanism is well-suited when trust relationships exist among domains, such as importing provider's reports into client's domain, or exchanging reports between arbitrary, peer-trusting domains (such as universities campuses). 

\subsection{Remote Enquiries}
\label{sec:disco:interdomain:remote}

The number of import/export channels that a DISco domain can maintain is necessarily limited, and since trust is not necessarily transitive, it makes little sense to provide mechanisms to relay reports of domain $P$ to domain $S$ through domain $D$. However, when an analyser in domain $S$ investigates suspicious traffic, information about the destination domain are more likely to be useful than knowing odds happening in the second or third relayer downstream $S$.

An option to achieve this is to encapsulate lookup requests for remote information and to relay them to a DISco-compatible system responsible of the intended domain. This mode of operation is heavily inspired and can be supported by the ``pull mode'' of i4 \cite{i4}, an information-exchange proposal between intrusion detection systems. Shared keys to establish the secure connection for the lookup request are generated iteratively and piggy-backed on BGP prefix advertisement\footnote{The major advantage of BGP-based key distribution is that key generation effort is limited and, compared to public key infrastructures, do not require a central certification authority. A domain $X$ that receives key $K_D$ for a given destination AS $D$ builds a derived key $K_D^Y = hash(K_D, Y)$ for every neighbour AS $Y$ it forwards the prefix advertisement.
It should be noted that other BGP systems between the enquirer and the enquired systems can impersonate the enquirer. This approach is thus better suited to ISP-to-remote-client enquiries than to support e.g. members of an overlay or partners of a grid computing.
}.

Again, the support for remote enquiries can be transparently added to a DISco system through the use of indirect stores and LOOKUP commands: a border system within the AS will scan BGP exchanges and install lookup indirections in the DISco store capturing the corresponding IP addresses, therefore playing the role of a DISco proxy. When a lookup request hits the proxy, it uses internally stored correspondence between IP blocks, AS path towards that block and the corresponding communication key.

Because there is no implied trust in this model -- only authenticated connection -- lookups should only enquiry about subjects for which $D$ can establish that motivation of the request is valid. Flows originating from or carried by the requester, for example, are good candidate subject, as the target system can easily verify involvement of the requester. Issuing generic requests about the load of routers or servers within the domain, however, would be denied.

\subsection{Activity Tracing}

With increasingly complex network technologies and ever rising use of near-real-time multimedia over best effort networks, even end-to-end control loops may need a clearer report of what happens within the network to optimise their behaviour and improve service availability for the end-user.

Two recent proposals, X-Trace \cite{xtrace} and NetReplay \cite{netreplay} investigate the opportunities of collecting network-generated monitoring at inter-domain scale and exploiting it to fine-tune the behaviour of end-to-end protocols, providing integration between service-level and network-level resilience. Similarly, as suggested in the Knowledge Plane \cite{863957}, suspicions of failure from end-systems can be incentive for the network-level resilience system to further investigate the current behaviour. We will focus on X-Trace in the remaining, as NetReplay is a problem-specific solution and that the knowledge plane is merely a conceptual proposal so far.

Upon reception of a packet carrying an X-Trace marker, protocol entities capture relevant part of their state and generate a report that is handled to a per-domain \emph{collector} and will eventually be logged in a \emph{report server} using the identity is mentioned in the X-Trace payload. This mechanism is generic and primarily aimed at debugging by human experts. Yet, we can extend it into a cross-layer and cross-domain query mechanism by having DISco lookups performed at the protocol entity and at the collector. The major limiting factor is the lack of a shared vocabulary between the requester and the replying system here, and the overhead due to systematic export policy checks.

It could also be tempting to import e.g. the presence of an X-Trace marker in a packet into DISco and have some of the analysers subscribe to such events in addition to reports generated by in-network probes, especially given that X-Trace-augmented packets carry a token that uniquely identifies the user-level ``activity'' that is under investigation by the end-system. When accessing a webmail, for instance, that very same token will appear in DNS requests, TCP connection establishments, and ideally, even internal web-to-IMAP requests\footnote{provided that end-systems and proxies are X-Trace-enabled}, providing precise correlation hints.

However, in the absence of an additional authentication of the emitter of the token and trust relationships between the emitter and the network resilience manager of a specific AS, it would be easy for a malevolent end-user to forge tokens in order to misguide the resilience mechanism of a victim network. Members of a DDoS botnet, for instance, could collectively use the same token, forcing the defence system to believe that something caused an unusual rate of retransmissions. Similarly, an intruder could introduce artificially different tokens to evade any detection based on token correlation.

\section{Integration with Existing Tools and Protocols}
\label{sec:disco:integration}

\subsection{Correlation Engines}
\label{sec:disco:integration:correlation}

Correlating events to draw conclusions is a key feature of the challenge detection process. We have specifically investigated the possibilities offered by ISS and Chronicles, two engines developed by projects partners and now present how they can be integrated with DISco.
More techniques certainly exist, and it is not intended to be exhaustive here, as different challenge might suit different correlation mechanisms. This speaks in favour of having correlation tasks kept out of DISco and implemented as part of the DISco ``clients''.

Information Sensing and Sharing framework (ISS \cite{iss}) has been developed at Lancaster as part of the functional composition framework of the ANA project\footnote{http://www.ana-project.org/}. While it wouldn't really be used inside DISco, it is a good way to build correlators and more sophisticated sensors on nodes. DISco's publish and subscribe system naturally extends the point-to-point data delivery that is already present in ISS.

Chronicles recognition is a mechanism for temporal events correlation that has been successfully applied to network intrusion detection at Orange Labs \cite{Morin03correlationof}. It features its own inter-connection mechanisms to build hierarchically organised distribution of detection and efforts, which again could benefit from DISco's peer-to-peer nature to improve scalability and dependability. Its strong dependency on time-related aspects puts an interesting constraint on how DISco could perform aggregation and filtering.

\subsection{Standard Network Monitoring Protocols}

Out of the existing network monitoring protocols, we have specifically investigated NetFlow, SNMP and syslog for interoperability with DISco. Both serve different purposes and are widely deployed in existing products. They typically fit a \emph{sensor} component of Fig.~\ref{fig:DISco-architecture}, but, as with external data stores discussed in Section~\ref{sec:disco:architecture:storage}, they need an additional \emph{translation} function that converts their reports into DISco event reports.

This translation typically includes the identification of event label as well as attributes extraction and conversion by matching the external notification against known patterns. In the case of NetFlow, the mapping can be pretty straightforward, especially thanks to the availability of compound values support in DISco. On the other hand, syslog entries would require a deep knowledge of the applications that generated them to proceed with matching and extraction, and, to a large extent, it would be preferable to alter the applications so that they natively support DISco pub/sub interfaces and have an external exporter translating DISco events into human-readable syslog events rather than the other way round.

Finally, the SNMP protocol provides much more than the functionality we propose in DISco, and its \emph{trap} mechanism (used to report notifications asynchronously) is the most interesting feature for our real-time approach. It should be noted, however, that despite an SNMP trap is linked to an object which value can be later looked up, SNMP daemons typically do not keep track of individual data evolution, and it wouldn't be possible to look into more details at a reported situation unless the aggregation happened after the translation step.
Again, thus, publishing translated SNMP traps into DISco should be seen as a cheap, transitional alternative to the native support of DISco in the network stack.

\section{Conclusion and Future Works}

In this paper, we have presented the design of an integrated event dissemination and storage system that meets the need of challenge detection system both in terms of real-time and bandwidth-limited notification and context lookups with variable granularity. The use of dynamic aggregation and filtering as events are forwarded is a fundamental feature of our solution, for which we propose an OMNet++ implementation based on the OverSim package. 

We conceptually illustrated the use of DISco abstractions (pub/sub, vocabularies and aggregation) on the scenario of identifying and tackling DDoS attacks and proposed guidelines to interconnect the distributed store of an autonomous system with external publishers and subscribers in order to assist service-level resilience managers.

The adaptive storage that we coupled with notification forwarding is intended to provide the necessary information for challenge diagnostic and remediation refinement. The most interesting research question in that regards include the strategies used by the retention manager to adjust information lifetime, the ranking of remediation strategies and the identification a-posteriori of context information that should guide the selection of weights for remediation mechanism activation.

Despite its desirable features, the performance of DISco -- and especially the ability of the event delivery subsystem to meet constraints of real-time challenge detection -- is still to be demonstrated. Our hope in that regards is that the additional path stretch caused by our peer-to-peer approach will be compensated by lighter processing load on aggregating and analysing systems (as opposed to a strictly hierarchical setup).

\section{Acknowledgements}

This work has been partially funded by EU project ResumeNet, FP7-224619. Sylvain Martin acknowledges the financial support of the Belgian National Fund of Scientific Research (FRS-FNRS).

\bibliography{disco-arxiv}

\newpage
\section{DISco API and Implementation Notes}
\label{sec:disco:api}
\label{DISco:omnet}

Our OMNet++ implementation provides services that allow components of the resilience framework to share information. Events can be \emph{published} and \emph{subscribed to}, as in any publish \& subscribe system, specifying at the same time some filters and / or aggregators. The identifiers for the events and their attributes are known throughout the system thanks to a vocabulary, as explained in section \ref{sec:disco:voc}. The aggregate event is \emph{delivered} to the subscriber through a callback mechanism. Finally, the subscriber can \emph{reply} to this event in order to provide feedback (such as annotation tags).

\paragraph{Publish (Template)}

\begin{tabular}{ll}
 \textit{Input:} & - event id\\
 & - template id\\
 & - list of attribute-type pairs
\end{tabular}

Before sharing any actual data, the publisher is required to send the template of the event format it uses. The template is associated to a \emph{template ID} that only has to be unique for the issuer. It then contains the list of (attributes ID, type) pairs, in the same order as these attributes will be present in any subsequent data messages. The attribute ID captures the semantic of the field (as defined by the VST) while the type captures the binary encoding and the available aggregation and comparison operations.

Format of events may be changed anytime by simply publishing a new template, using a different \emph{template ID}.

\paragraph{Publish (Data)}

\begin{tabular}{ll}
 \textit{Input:} & - event id\\
 & - template id\\
 & - list of attribute values
\end{tabular}

A published event is only composed of the event identifier, the identifier of the template it uses, and the list of attribute values following that template. The publisher remains a simple process that sends information to the system without consideration to how (and how much) aggregation or filtering has to be performed on these events before being transmitted to the subscribers (if any is interested in its publications).

The corresponding template is only required once, but it must be issued before any data using this template is published. This is indeed necessary to perform filtering and aggregation during the forwarding down the multicast tree.

\paragraph{Subscribe}

\begin{tabular}{ll}
 \textit{Input:} & - event id (possibly with wildcard) \\
 & - constraints on attributes (filters) \\
 & - attributes to discard \\
 & - specification of aggregation
\end{tabular}

When interested in a particular type of events, the subscriber provides the event identifier and specifies the desired filtering, aggregation type and granularity of received events. Thanks to the hierarchical setup of the events, it is possible to simultaneously subscribe to a whole set of similar events by using wildcards at the end of the name. When translating the event name to a numerical ID (using the VST), a wildcard is represented by a mask in a very similar way to what is done with IP subnets. 

Filters can be used to associate constraints on attributes, such that only events with matching values will be taken into account and included in the aggregate to be forwarded to the subscriber. The most obvious kind of constraints that can be defined are constraints on the range of an attribute value, specifying either a lower bound, upper bound, or both. For other \emph{types} of attributes, exact match of the value may be pertinent. 

Then, the subscriber may list attributes (considered by the subscriber of zero interest) to be discarded, such that they will not be included in the forwarded aggregate. This allows more flexibility than having to list explicitly all the desired attributes. Since a subscriber will usually not know all the available attributes before receiving the template (thus after the first subscription), it is the expected behaviour for subscribers to perform several subscribe requests for a single event ID, but with varying specifications, especially regarding the list of discarded attributes.

Together with these filters, aggregators can be appointed. Subscribers may specify the operation to be applied (based on pre-defined ones) on a per-attribute basis. The actually executed code will depend on both the operation (selected by the subscriber) and the attribute type (defined in the template). Default operations are provided for attributes not mentioned by the subscriber, based on their types.

The last input element is the desired granularity level. It can be specified both in terms of maximum aggregation level (i.e., no more than $x$ elementary events in an aggregated one) or publication rate (i.e., no more than $x$ aggregated events per second). These are complementary and can be specified at the same time, in which case the aggregate will be forwarded as soon as one of those two requirements triggers.

\paragraph{Deliver callback (Template)}

\begin{tabular}{ll}
 \textit{Output:} & - event id\\
 & - template id\\
 & - list of attribute-type pairs
\end{tabular}

Before receiving its first event, a subscriber will be delivered the associated template. The message is very similar to its publish counterpart, but the template itself will most probably be different of the published one, due to aggregation and discarded attributes.

\paragraph{Deliver callback (Data)}

\begin{tabular}{ll}
 \textit{Input:} & - event id\\
 & - template id\\
 & - list of attribute values\\
 & - number of aggregated events\\
 & - aggregation period
\end{tabular}

The delivered message naturally contains the \emph{event id}, \emph{template id} and the actual data, but also context information about the aggregation, namely the number of aggregated events and the aggregation period. Even if these two pieces of information are not actually used by the subscriber, they are required during the forwarding process in order to be able to aggregate events at several stages in the multicast tree.

\paragraph{Reply}

\begin{tabular}{ll}
 \textit{Input:} & - event id\\
 & - constraints on attributes \\
 & - constraints on timestamps \\
 & - tags
\end{tabular}

Replies allow a subscriber to send back information to the system following a received event. The main purpose of this service is to provide an easy mechanism to annotate events. These annotations (tags) are useful to keep trace of correlation steps and to indicate relevance of events in order to adjust their lifetime in the store. A single reply can be used to annotate a group of similar events based on timestamps and optional attribute constraints. The encoding of constraints is similar to the one used in the \texttt{subscribe} call.

In order to support the publication replies, which potentially have to ``revert'' aggregation from multiple sources, we suggest to reuse the mechanism of \emph{z-Filters} presented in \cite{lipsin}. Every link between two members of the pub-sub system is associated with a identifier $L_i$ that translates into a link filter $H(L_i)$ where $H$ is a bloom-compatible hash function. Every time an event is sent over a link, the link identifier is OR-ed with the z-Filter contained in the packet. Similarly, when a collection of messages need to be aggregated, the z-Filter of the aggregated message is the OR superposition of the initial messages. As a result, the message a subscriber receives contains the superposed identities $Z$ of all the links used between himself and the sources of information he receives. Consequently, if a \emph{reply} message contains $Z$, nodes can forward it backward by AND-ing each of their $L_i$ with $Z$ and forwarding only on links that trigger a match. As in every application of bloom filters, proper selection of mask width and amount of bit set allows one to tune the rate of false positive (a message being forwarded on a link it shouldn't take).

\end{document}

%% file: FigMacros.tex
\usepackage{graphicx}
\usepackage{ifthen}
\usepackage{listings}
\usepackage{multicol}

\graphicspath{{./figures/}}

\newcommand{\fig}[3][1]{
  \begin{figure}[tb]
    \begin{center}
      \includegraphics[width=#1\columnwidth]{#2.pdf}
      \caption{#3}
      \label{fig:#2}
    \end{center}
  \end{figure}
}

